\documentclass[a4paper,11pt]{article}
\usepackage{graphics}
\usepackage{jheppub}
\usepackage{epsfig}
\usepackage{subfig}
\usepackage{float}
\usepackage{slashed}
\usepackage{soul}
\usepackage{cancel}
\usepackage{multirow}
\usepackage{amsmath,amssymb,mathrsfs}
\usepackage[utf8]{inputenc}
\usepackage{color}
\definecolor{ourcolor}{rgb}{0.7, 0.25, 0.05}

\def\be{\begin{equation}}
\def\ee{\end{equation}}

\def\bea{\begin{eqnarray}}
\def\eea{\end{eqnarray}}

\title{\color{ourcolor}{Leptoquarks: 750 GeV Diphoton Resonance and IceCube Events}}

\author[]{Ujjal Kumar Dey,}
\author[]{Subhendra Mohanty}
\author[]{and Gaurav Tomar}
\affiliation[]{Physical Research Laboratory, Ahmedabad 380009, India.}

\emailAdd{ujjaldey@prl.res.in}
\emailAdd{mohanty@prl.res.in}
\emailAdd{tomar@prl.res.in}

\abstract{The recent data from ATLAS and CMS hint at a new resonance at around 750 GeV in the diphoton invariant mass distribution. The  explanation of the significantly large cross section for this diphoton resonance requires coloured particles in its loop-induced production via gluon fusion and subsequent decay to diphoton. A natural candidate for the coloured particle is the colour-triplet leptoquark, lying in the mass range 375-1000 GeV, which can account for such large cross section. The leptoquarks in this mass range can also be produced resonantly from neutrino and quark interactions at IceCube and provide an explanation to the PeV-energy neutrino events. In this work, we show that the scalar leptoquark with quantum number $(3,2,7/6)$ can uniquely provide a unified explanation to both the PeV IceCube events and the 750 GeV diphoton resonance.}


\allowdisplaybreaks

\begin{document}
\maketitle

\section{Introduction}
\label{sec:intro}
The alluring hint of a new resonance, responsible for an excess in the diphoton invariant mass around 750 GeV, might be the harbinger of the long sought-after new physics (NP) beyond the standard model~\cite{Aaboud:2016tru, Khachatryan:2016hje}. The advent of this signal is embraced by the particle physics community with sheer excitement which leads to a host of phenomenological models\footnote{See the recent review on 750 GeV digamma excess and the references therein~\cite{Strumia2016}. Also see~\cite{Staub2016} where a number of models are checked using the software package SARAH.}. The number of these models will be narrowed down by correlating them with other experimental observations as well as by forthcoming data from the LHC. The Landau-Yang theorem dictates that the final state diphoton can come from a spin-0 or spin-2 particle. We consider the spin-0 possibility and denote it by $\Phi$. To have a significant cross section for the process $pp\to \Phi \to \gamma \gamma$ the scalar resonance $\Phi$ must be produced from $gg$ or $qq$ vertices and decay to diphoton via exotic fermions, gauge bosons or scalars in the loop. 
One of the well-motivated candidates for these loop particles is leptoquark (LQ) which simultaneously carries colour and $SU(2)_{L}\otimes U(1)_Y$ quantum numbers. A comprehensive study of LQs has been performed in \cite{Buchmuller1987, Dorsner2016}. The LQ models have been used to explain the 750 GeV diphoton resonance~\cite{Bauer2015,Murphy2015}. LQs can be both of vector and scalar nature. Due to their unusual quantum numbers LQs can couple quarks and leptons. This is why a natural arena for testing the LQ models of diphoton resonance is at IceCube.
South-pole based neutrino detector IceCube has witnessed the highest energy neutrino events till date. It has observed 54 ultra high energy (UHE) neutrino events which is spread in the energy range from TeV to PeV in its four years of data taking \cite{Aartsen:2015zva}. The extraterrestrial nature of these events is also confirmed at more than 6$\sigma$ confidence level~\cite{Aartsen:2015zva}. In IceCube data there is a problem of explaining the non-observation of Glashow resonance at 6.3 PeV. This tension can be somewhat alleviated by taking a steeply falling power-law neutrino flux. Even with the steep power-law flux the three highest energy (PeV) events can not be accounted for.
It has been shown that near-TeV range LQs can significantly contribute in the UHE neutrino events~\cite{Anchordoqui:2006wc, Alikhanov2013, Barger2013, Dey2016, Dutta:2015dka}. LQs can be produced on-shell from highly energetic incoming neutrinos scattering with nucleons in the ice molecules, and then its subsequent decay can enhance the high energy shower event rates at the IceCube.
In this paper we propose a common explanation to the 750 GeV diphoton resonance and the PeV IceCube events using a scalar LQ model. Starting from a general set of scalar LQs we test each of the LQs against the observed diphoton signal and then zeroing on the best-suited LQ we check its competence in explaining the IceCube events. Previously a dark matter model connecting the diphoton excess and the IceCube signal has been explored in~\cite{Morgante2016}.
This paper is organized as follows. In the next Sec.~\ref{sec:model} we describe the relevant interactions among the 750 GeV scalar $\Phi$ and LQs as well as the important properties of the LQs. In Sec.~\ref{sec:diph} we discuss the performance of various scalar LQs with respect to the diphoton signal. The analysis IceCube events in the context of the present LQ scenario is given in Sec.~\ref{sec:IC}. Finally we summarise and conclude in Sec.~\ref{sec:concl}
\section{Model}
\label{sec:model}
We augment the standard model (SM) with one real singlet scalar field $\Phi$ (which we identify as the 750 GeV resonance) and the scalar leptoquark $\eta$. The relevant Lagrangian is given by, 
\begin{align}
\mathcal{L} = \mathcal{L}_{SM} + \frac{1}{2} m^2_{\Phi}\Phi^2 + (\lambda_s \Phi + m_{LQ}^2)|\eta|^2,
\label{eq:lag}
\end{align}
where $m_{\Phi}$ is the mass of the scalar $\Phi$, $\lambda_s$ is the dimensionful coupling between the leptoquark and $\Phi$, and $m_{LQ}$ is the mass of the LQ. The singlet scalar $\Phi$, in principle, can have mixing with the SM Higgs ($H$) via terms like $\Phi^{\dagger}\Phi H^{\dagger}H$ and also if $\Phi$ gets a vacuum expectation value (vev) there exists the possibility of interactions which may lead to the decay of $\Phi$ to two SM Higgs. In our analysis we assume such interactions to be negligible. A few words about the dimensionful coupling $\lambda_{s}$ are also in order. Note that the $\lambda_{s}$ can lead to a correction to the mass of the singlet scalar and this would be of the order of $(1/16\pi^{2})\lambda_{s}^{2}(\Lambda^{4}/m_{LQ}^{4})$, where $\Lambda$ is the cut-off of the theory. A conservative choice of $\Lambda \sim 1$ TeV would result in a $\lambda_{s} \sim $ 6-7 TeV. In Table~\ref{tab1} we enlist the scalar LQs (the generic $\eta$ of Eq.~\ref{eq:lag}), with their corresponding quantum numbers, and mention which of them give(s) rise to tree-level operators leading to rapid proton decay~\cite{Dorsner2016}. These are the only possible set of scalar LQs which can be relevant for the explanation of IceCube data as they couple with SM neutrinos and quarks.
\begin{table}[!htbp]
\centering
\begin{tabular}{|c|c|c|}
\hline
Leptoquarks     & $SU(3)_{c}\otimes SU(2)_{L}\otimes U(1)_{Y}$ & \begin{tabular}[c]{@{}c@{}}Proton decay\\ (at tree-level)\end{tabular} \\ \hline \hline
$S_{3}$         & ($\bar{3},3,1/3$)                            & Yes                                                                    \\ \hline
$R_{2}$         & ($3,2,7/6$)                                  & No                                                                     \\ \hline
$\widetilde{R}_{2}$ & ($3,2,1/6$)                                  & No                                                                     \\ \hline
$S_{1}$         & ($\bar{3},1,1/3$)                            & Yes                                                                    \\ \hline
\end{tabular}
\caption{List of the scalar leptoquarks with corresponding gauge quantum numbers.}
\label{tab1}
\end{table}

At this point it is worth-mentioning that even though the leptoquarks $R_{2}$ and $\widetilde{R}_{2}$ does not have tree-level diquark couplings which give rise to rapid proton decay, there may exist dimension five operators which can lead to high baryon number violation resulting in rapid proton decay. But these higher dimension operators can be forbidden in a systematic fashion~\cite{Arnold:2013cva}. 
In passing we also note the existing bounds, from the LHC data, on the masses of LQs. According to the CMS data the first generation scalar leptoquarks with mass less than 1010 (850) GeV are excluded for the branching fraction $\beta=1.0~(0.5)$ into lepton and quark \cite{Khachatryan2016}. For the second generation leptoquark with mass less than 1080 (760) GeV is excluded \cite{Khachatryan2016}. The ATLAS collaboration, on the other hand, puts leptoquark mass bounds at 95\% confidence level to be 1100 GeV and 1050 GeV (1160 GeV and 1040 GeV) for first and second generation leptoquarks, respectively, assuming a 100\% branching ratio into a charged lepton and a quark~\cite{Aaboud2016}. The mass bounds on the vector leptoquarks are much severe in comparison to the scalar leptoquarks \cite{Khachatryan2016}. These bounds can be evaded by LQs which link the first generation quarks with third generation leptons. We will consider these types of scalar LQs in the analysis performed in this paper.

\section{Diphoton signal}
\label{sec:diph}
Clearly the production of $\Phi$ can take place through LQ loops, which also promotes its decay into $\gamma\gamma$. The other possible decay channels of $\Phi$ are $gg$, $WW$, $\gamma Z$ and $ZZ$, depending on the quantum numbers of the LQs. Also, if $m_{\Phi} < 2m_{LQ}$ the tree-level decays of $\Phi$ to the LQs are kinematically forbidden; if this condition is not satisfied then the tree-level decay of $\Phi$ to two leptoquarks will dominate and the required diphoton signal strength can not be obtained. Thus the lower end value of the mass of the leptoquark should be 375 GeV to conform with the observed diphoton signal.
The decay widths of $\Phi$ into $gg$ and $\gamma\gamma$ channels are given by,
\begin{subequations}
\begin{align}
\Gamma (\Phi \to \gamma \gamma) =& \frac{\alpha^2 N^2_c N^2_f}{1024\pi^3}\frac{m^3_\Phi}{m_{LQ}^2}
                        \left|\frac{\lambda_s}{m_{LQ}}\left(\sum_i Q^2_i\right)A_0\left(\frac{m^2_\Phi}{4m_{LQ}^2}\right)\right|^2,\\
\Gamma (\Phi \to gg) =& \frac{\alpha_s^2 N^2_f}{512\pi^3}\frac{m^3_\Phi}{m_{LQ}^2}
                        \left|\frac{\lambda_s}{m_{LQ}} A_0\left(\frac{m^2_\Phi}{4m_{LQ}^2}\right)\right|^2,
\end{align}
 \label{rdw}
\end{subequations}
where the form factor $A_0$ is given as,
\begin{align}
A_0(\tau) = -\frac{1}{\tau^2}(\tau - f(\tau)),~\mbox{with}~\tau = \frac{m^2_\Phi}{4m_{LQ}^2},
\end{align}
and
\[
    f(x)= 
\begin{cases}
    \left[\mbox{sin}^{-1}(\tau)\right]^2,& \text{if } \tau \leq 1,\\
    -\frac{1}{4} \left[\mbox{log}\frac{1+\sqrt{1-\tau^{-1}}}{1-\sqrt{1-\tau^{-1}}}-i\pi\right]^2, & \text{if } \tau > 1.
\end{cases}
\]
In above equations the quantity $N_{f}$ can be thought of as the number of flavours of the leptoquarks and we consider the most economical case with $N_{f} = 1$. The ATLAS \cite{Aaboud:2016tru} and CMS \cite{Khachatryan:2016hje} data suggest that the cross section $\sigma\in [3,13]$ fb. In our analysis we use this range and calculate the allowed parameter space. First, we check the consistency of the viable scalar leptoquarks against the diphoton signal and then go on to use the best suited LQ in our IceCube analysis.
As mentioned before, in our model, the production of $\Phi$ takes place mainly through gluon fusion via scalar leptoquark loops, the subsequent decay into two photons also takes place via similar loops.
The diphoton signal cross section is given by,
\begin{align}
\sigma(pp\to \Phi \to \gamma \gamma) = \frac{1}{m_{\Phi} s\Gamma_{\rm tot}}
         [c_{gg} \Gamma(\Phi \to gg)]
         \Gamma(\Phi \to \gamma \gamma),
         \label{meq}
\end{align}
where $\Gamma_{\rm tot}$ is the total decay width of $\Phi$. Here the parton integral $c_{gg}$ is given by,
\begin{align}
c_{gg} = \frac{\pi^{2}}{8}\int_{m_{\Phi}^{2}/s}^{1}
         \frac{dz}{z}g(z)g\left(\frac{m_{\Phi}^{2}}{zs}\right).
\end{align}
The numerical value of $c_{gg}$ is $\sim 2137$ \cite{Franceschini2015} for $m_{\Phi}=750$ GeV at $\sqrt{s} = 13$ TeV. The ATLAS and CMS collaborations suggest $\sigma$ to be lying in between 3-13 fb. 

As we point out before, in the connection to the IceCube events, there remain only four leptoquarks (see Table~\ref{tab1}) which can give a common explanation to diphoton signal and IceCube events. Below we test their merits in explaining the diphoton signal.

For the $S_3$ leptoquark there are three possible states with charges $Q_1=4/3,~Q_2=-2/3$, and $Q_3=1/3$ which give contributions to the decay width $\Gamma (\Phi \to \gamma\gamma)$. We calculate the decay widths $\Gamma (\Phi \to \gamma\gamma)$ and $\Gamma (\Phi \to gg)$ using Eq. \ref{rdw} by taking $N_f=1$. We use this information to calculate $\sigma$ from Eq. \ref{meq}, which should lie in the range of 3-13 fb. We find out that for the minimal scenario $(N_f=1)$, it is possible to accommodate the  observed diphoton signal for $S_3$ leptoquark with its mass between 375-740 GeV.

The possible charge states for $\tilde R_2$ is $Q_1=2/3$ and $Q_2=-1/3$. We take them into account to calculate $\Gamma (\Phi \to \gamma\gamma)$ taking $N_f=1$. We also evaluate $\Gamma (\Phi \to gg)$ for the further calculation of $\sigma$ which should lie in the range reported by ATLAS and CMS collaborations. We find that this leptoquark can also be a good candidate for the explanation of diphoton anomaly if its mass lies between 375-500 GeV in the minimal scenario $(N_f=1)$.

In the connection to IceCube PeV events, the third possibility is $S_1$ leptoquark with the only possible charge $Q=1/3$. We compute the decay widths into photons and gluons channels, and finally calculate $\sigma$ for $N_f=1$. We find that for $\sigma\in [3,13]$ fb, the possible mass range of $S_1$ leptoquark is 375-380 GeV. 

\begin{figure}[!htbp]
\centering
\subfloat[\label{fig:r2dijet}]{
\includegraphics[scale=0.66]{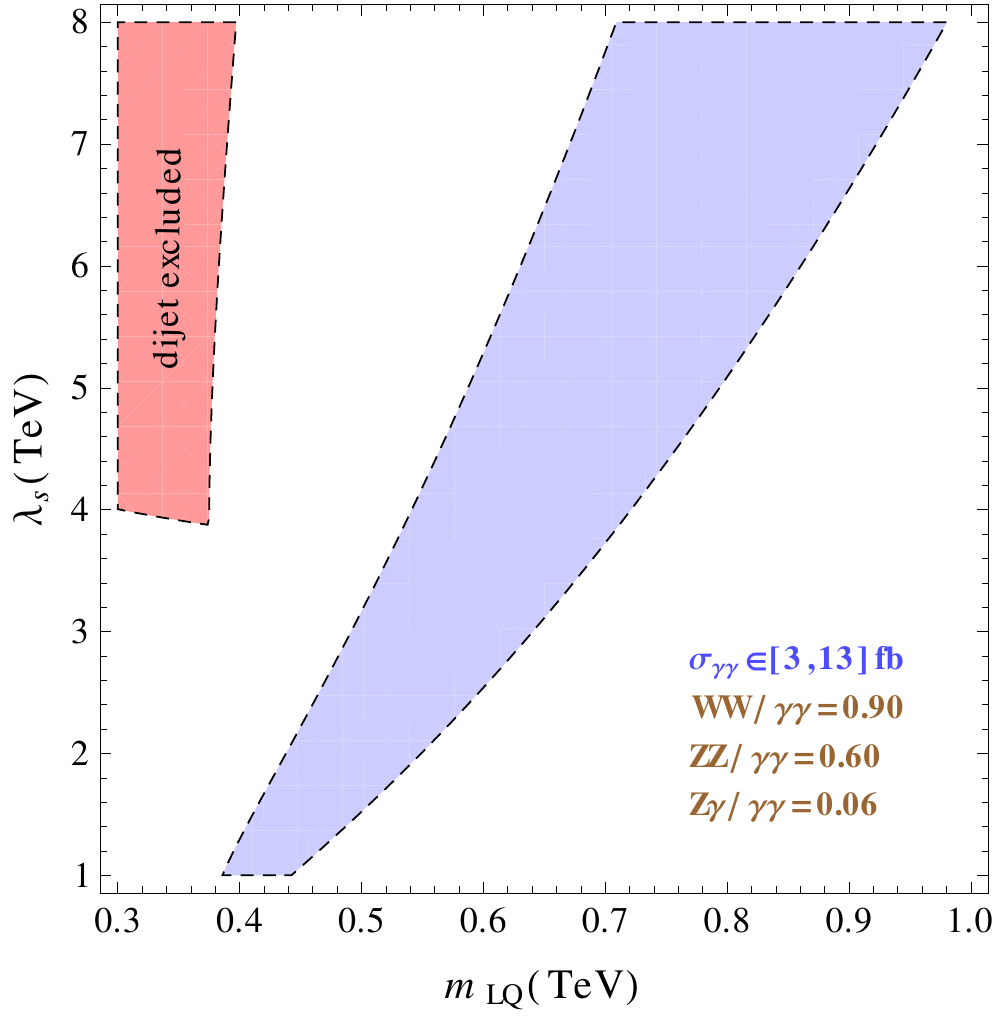}
}~~~~
\subfloat[\label{fig:r2sigmlq}]{
\includegraphics[scale=0.71]{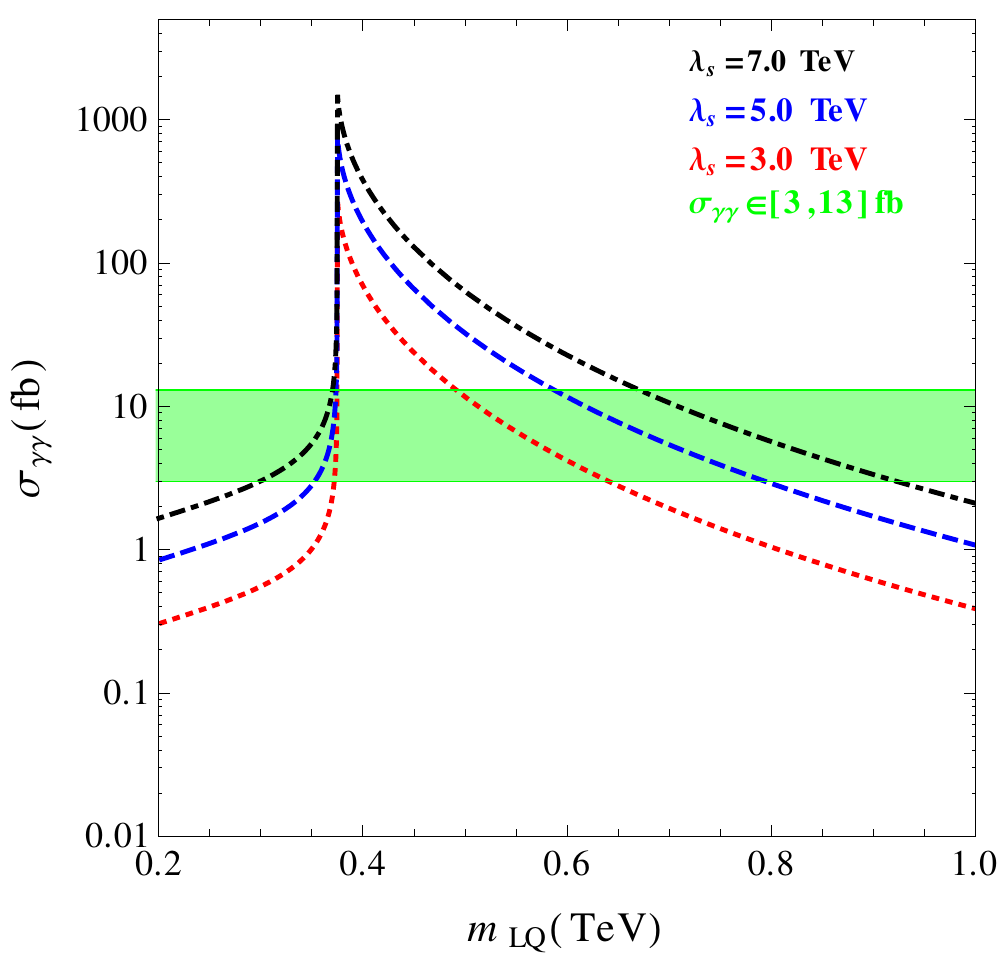}
	  }
\caption{(a) The diphoton rate as a function of $m_{LQ}$ and $\lambda_s$ for the $R_2$ leptoquark taking $N_f=1$. (b) The diphoton signal cross section as a function of leptoquark mass $m_{LQ}$ for different values of the dimensionful coupling $\lambda_s$.}
\label{fig:r2plots}
\end{figure}

In the case of $R_2$ leptoquark, the possible states are of charges $Q_1=2/3,~Q_2=5/3$. We evaluate the decay widths into photons and gluons channels and calculate $\sigma$ taking $N_f=1$. We plot the allowed cross section $\sigma\in [3,13]$ fb in Fig. \ref{fig:r2dijet} varying the coupling $\lambda_s$ and leptoquark mass $m_{LQ}$. Clearly for $R_2$ leptoquark also, it is possible to accommodate the  observed diphoton signal for its mass between 375-1000 GeV. We show the diphoton cross section as a function of leptoquark mass in Fig. \ref{fig:r2sigmlq} which indicates that for different values of $\lambda_s$, it is possible to explain the diphoton anomaly easily. For $m_{LQ}<375$ GeV, $\Phi$ can decay into $R_2$ leptoquark at tree level and thus reducing the branching fraction to diphoton, which in turn reduces the total cross section as can be seen in Fig. \ref{fig:r2sigmlq}.

\begin{table}[]
\centering
\begin{tabular}{|c|c|c|c|c|}
\hline
Leptoquarks               & $\gamma Z/\gamma\gamma$ & $WW/\gamma\gamma$  & $ZZ/\gamma\gamma$ & $gg/\gamma\gamma$\\ \hline \hline
$S_3$                     & 4.40                     & 3.45              & 8.37              & 9.24\\ \hline
$R_2$                     & 0.06                     & 0.90               & 0.60              & 4.84\\ \hline
$\widetilde R_2$              & 5.02                     & 30.33              & 9.14              & 1.6 $\times 10^2$\\ \hline
$S_1$                     & 0.60                     & 0.0                & 0.09              & 4.0 $\times 10^3$ \\ \hline
\end{tabular}
\caption{The leading order branching fractions of $\Phi$ decay into different final states through considered leptoquarks.}
\label{tab2}
\end{table}

In Table \ref{tab2}, we have given the branching fractions of $\Phi$ decay into different final states with respect to the $\gamma\gamma$ final state for various choices of LQs. It is clear that the scalar leptoquarks $S_3$ and $\tilde R_2$ have a large brancing fractions into $\gamma Z,~WW$, and $ZZ$ channels in comparison to $\gamma\gamma$. So these leptoquarks are in conflict with the observations, where no excess has been found. So there remain only two possibilities, $R_2$ and $S_1$ which can explain the diphoton anomaly while remaining consistent with existing data. Another stringent constraint on these leptoquarks will come from the dijet data~\cite{Khachatryan2016a}. In Fig. \ref{fig:r2dijet}, we have shown the dijet bounds on the allowed parameter space for the LQ $R_{2}$. Clearly the dijet bounds are not strong enough to constrain the allowed parameter space. However, we have checked that for $S_{1}$ the diphoton allowed region is excluded by the dijet constraints. Moreover, from Table \ref{tab1} we see that $S_1$ gives rise to proton decay at tree level. Thus the best possible LQ candidate emerging from our discussion so far would be $R_{2}$ and we use it for the explanation of the IceCube events. It is worth-mentioning that the LQ, $R_{2}$, can explain the observed ratios $R(D)$ and $R(D^{\ast})$ of $B$-meson decay. Apart from this, along with other flavour sector predictions and muon ($g-2$) implications of $R_{2}$ are explored in~\cite{Dorsner2013}.  
\section{IceCube events}
\label{sec:IC}

Four years IceCube data from 2010 to 2014 for a total livetime of 1347 days observed a total of 54 events. Earlier 3-yr data tabulated 37 events which consist of 9 track and 28 shower events~\cite{Aartsen:2014gkd}. The fourth year data enlists 17 more events; 6 track and 11 shower events, but none of them exceeds 1 PeV~\cite{Aartsen:2015zva}. Also it is worth mentioning that the track events arise form the charged-current interactions of $\nu_{\mu}$s whereas the shower (\textit{aka} cascade) events originate from the charged-current interactions of $\nu_{e}$ and $\nu_{\tau}$ and neutral-current interactions of all flavours of neutrinos. The three highest energy events are all shower events.  
Under the assumption of isotropic astrophysical neutrino flux consisting of equal flavours at the Earth, the all-flavour spectrum with neutrino energies between 25 GeV and 2.8 PeV is depicted by the best-fit flux~\cite{Aartsen:2015knd}, 
\begin{align}
E_{\nu}^{2}\Phi(E_{\nu}) = \left(6.7^{+1.1}_{-1.2}\right)
         \times 10^{-8}\left(\frac{E_{\nu}}{100\text{TeV}}\right)^{-0.50\pm 0.09}~\text{GeV}\text{cm}^{-2}\text{s}^{-1}
\text{sr}^{-1}~.
\end{align} 
For our analysis also we consider the standard (1:1:1) flavour ratio on Earth and use \texttt{CTEQ6l} leading order PDF sets~\cite{Pumplin:2002vw} for the cross section calculations.
The expected number of events in a given deposited energy bin at IceCube is given by,
\begin{align}
\mathcal{N} = 
n_{T}T\int_{E_{\rm{min}}^{\rm{bin}}}^{E_{\rm{max}}^{\rm{bin}}} 
dE \left(\int_{E}^{\infty}dE_{\nu}
\left[\frac{d\sigma_{\rm{NC}}}{dE} + 
\frac{d\sigma_{\rm{CC}}}{dE}\right]
\Phi(E_{\nu})\Omega(E_{\nu})\right),
\end{align} 

where $n_{T}$ is the effective number of target nucleons at IceCube, and can be approximated as $6.0\times 10^{38}$; $T$ represents the time of exposure which is 1347 days; $d\sigma/dE = (1/E_{\nu})(d\sigma(E_{\nu})/dy)$ and $E$ is the deposited visible energy, $y$ being the visible daughter lepton energy fraction~\cite{Dutta:2015dka}. The effective solid angle coverage is taken care of by the quantity $\Omega(E_{\nu})$ which encapsulates the shadowing effect of the neutrino coming from the northern hemisphere~\cite{Gandhi:1995tf, Gandhi:1998ri}. 
Previously the explanation of IceCube events are given from various physical scenarios e.g., decaying dark matter or general dark matter phenomenology~\cite{Esmaili2013, Esmaili2014, Bhattacharya2015, Murase2015, Dev2016, Fiorentin2016, DiBari2016}, resonantly produced LQs~\cite{Alikhanov2013, Barger2013, Dey2016, Dutta:2015dka}, $R$-parity violating supersymmetric models~\cite{Dev2016a}, Lorentz invariance violation~\cite{Tomar2015}. A few general aspects (e.g., flavour composition of neutrinos, flux of incoming neutrinos etc.) of IceCube events can be found in~\cite{Mena2014, Chatterjee2014, Palomares-Ruiz2015, Vincent2016, Kistler2016}.
As we mention in the previous section, we would like to test the applicability of $R_{2}$ as the explanation of IceCube events. The relevant interactions of $R_{2}$ with SM fermions are given by the following Lagrangian~\cite{Dorsner2016},
\begin{align}
 \mathcal{L} &= Y_{1}^{ij} \bar u_{iR} (R_{2}^{T}i \tau_{2} L_{jL})
              + Y_{2}^{ij} \bar \ell_{iR} R_2^\dagger Q_{jL}  
              + \mbox{h.c.} \notag \\
             &=  Y_{1}^{ij} \bar{u}_{iR}l_{jL}R_{2}^{5/3} + 
                 Y_{1}^{ij} \bar{u}_{iR}\nu_{jL}R_{2}^{2/3} +
                 Y_{2}^{ij} \bar{l}_{iR}u_{jL}R_{2}^{5/3\ast} +
                 Y_{2}^{ij} \bar{l}_{iR}d_{jL}R_{2}^{2/3\ast} +
                 \mbox{h.c.}
\end{align}
where $Y_{1}$ and $Y_{2}$ are complex $3\times 3$ Yukawa type matrices. The $R_2$ leptoquark has two possible states, $R^{5/3}_2$ and $R^{2/3}_2$, with electric charges $5/3$ and $2/3$ respectively. Clearly among these two states the relevant one for IceCube will be the state $R^{2/3}_2$ which can give rise to processes like,
\begin{subequations}
\begin{align}
&\nu u \to  R^{2/3}_2 \to \nu u ~~ (\rm{NC ~type}) \\
&\nu u \to  R^{2/3}_2 \to l^{+} d ~~ (\rm{CC ~type})~.
\end{align}
\label{e:nuproc}
\end{subequations}
Evidently the first kind of processes will involve only $Y_{1}^{ij}$ couplings whereas the second type of processes will involve the product \textit{e.g.,} $Y_{1}^{ij}Y_{2}^{kl}$. Also the incoming neutrino with PeV energy will have sufficient centre-of-mass energy ($\sqrt{2m_{p}E_{\nu}} \geq m_{LQ}$) to result in a resonant $s$-channel LQ exchange. 
\begin{figure}[!htbp]
\begin{center}
\includegraphics[scale=0.6]{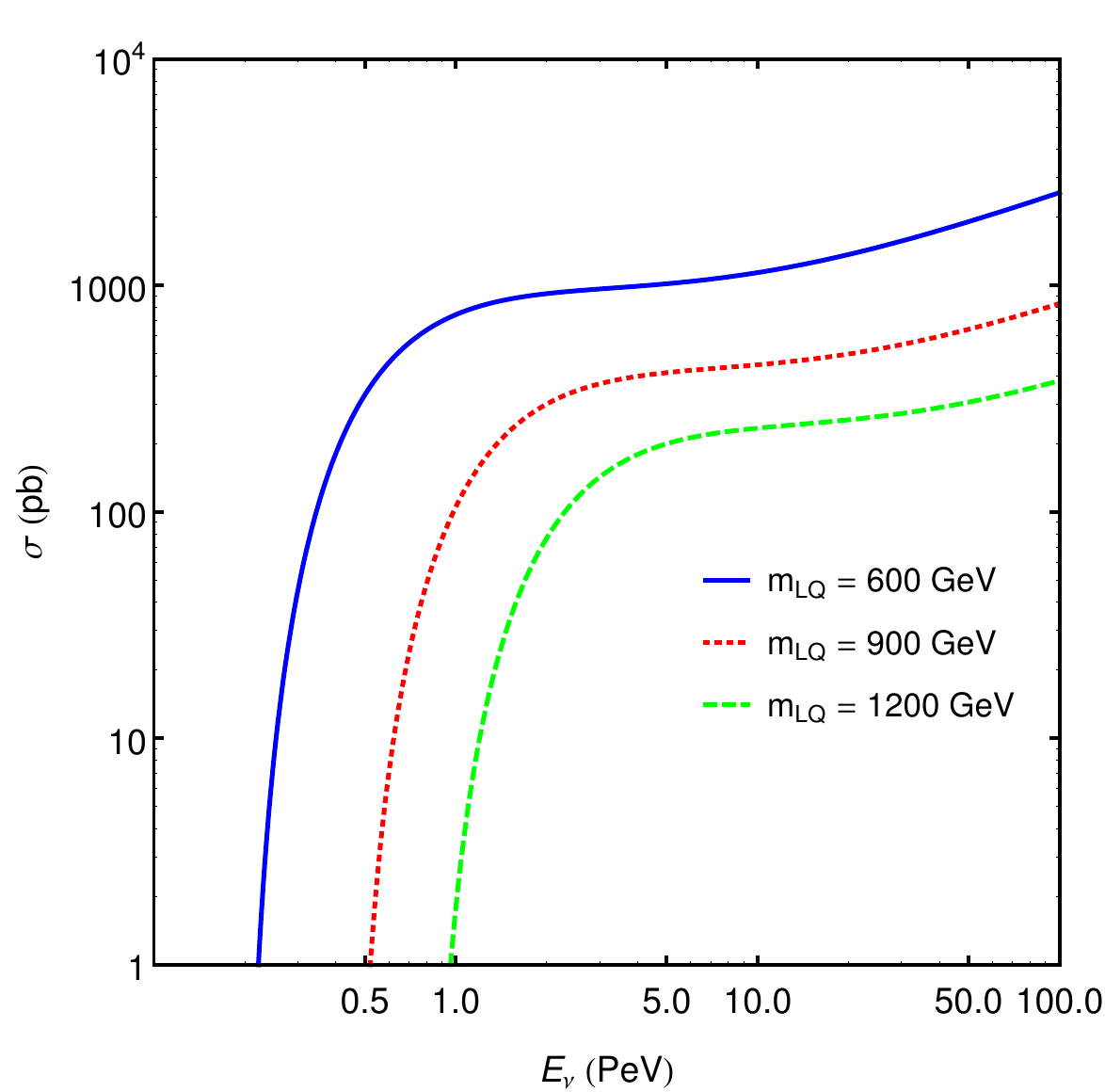}
   \caption{Neutrino-nucleon scattering cross section for various LQ masses and unit couplings.}
\label{f:lqsecn}
\end{center}
\end{figure}
The differential resonance cross sections for the above mentioned processes can be written, in a generic form, as~\cite{Anchordoqui:2006wc, Dutta:2015dka},
\begin{align}
\frac{d\sigma_{\rm{NC/CC}}}{dy} = \frac{\pi}{2}
              \mathscr{R}_{\rm{NC/CC}}
              \frac{\mathcal{U}(m_{LQ}^{2}/s)}{s},
\end{align}
where $y$ is the daughter lepton energy fraction ($\equiv E_{\ell}/E_{\nu}$); $\mathcal{U}$ represents the parton distribution function for up-type quarks and the factor $\mathscr{R}$ can be written as,
\begin{align}
\mathscr{R}_{\rm{NC}} = \frac{(Y_{1}^{ij})^{4}}
                         {(Y_{1}^{ij})^{2} + (Y_{2}^{kl})^{2}}~~
        \mbox{and} ~~
\mathscr{R}_{\rm{CC}} = \frac{(Y_{1}^{ij}Y_{2}^{kl})^{2}}
                         {(Y_{1}^{ij})^{2} + (Y_{2}^{kl})^{2}}~.
\end{align}
In Fig.~\ref{f:lqsecn} we show the neutrino-nucleon cross section, which occurs via LQ, for various masses of LQ where we assumed unit couplings to calculate this cross section.
At this point it is worth mentioning that to evade the existing mass bounds on the scalar LQs~\cite{Khachatryan:2014ura, Aad:2015caa, Khachatryan2016} we will consider the LQs which will have couplings with third generation leptons and first generation quarks. Thus in our representative Eq.~\ref{e:nuproc} the neutrinos are $\nu_{\tau}$ and $u$ will be the up-quark and as a matter of fact the relevant Yukawa type couplings will be $Y_{1}^{13}$ and $Y_{2}^{31}$.

\begin{figure}[!htbp]
\centering
  \subfloat[\label{f:mlq780}]{
\includegraphics[angle=-90,scale=0.28]{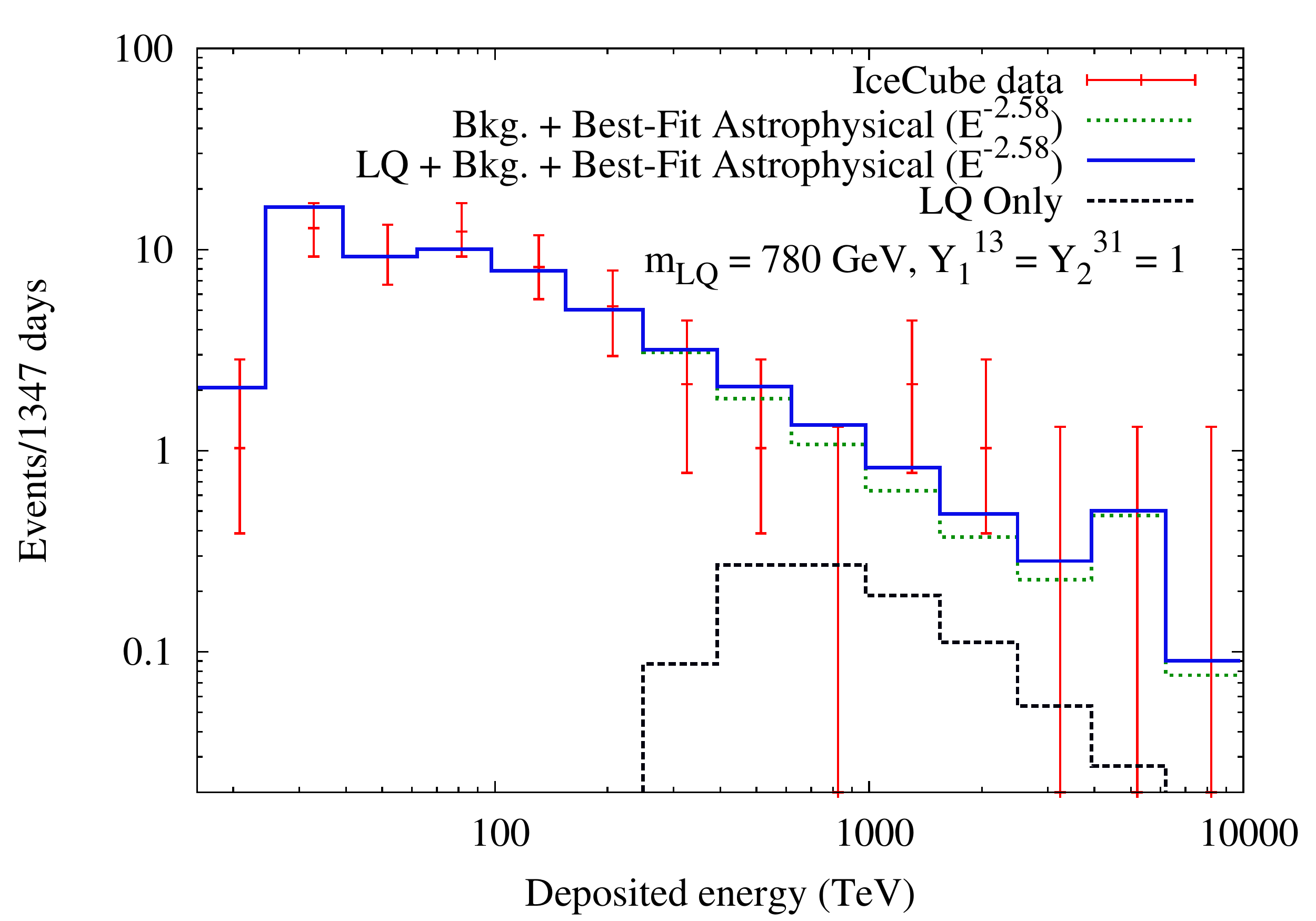}
   }
  \subfloat[\label{f:mlq900}]{
\includegraphics[angle=-90,scale=0.28]{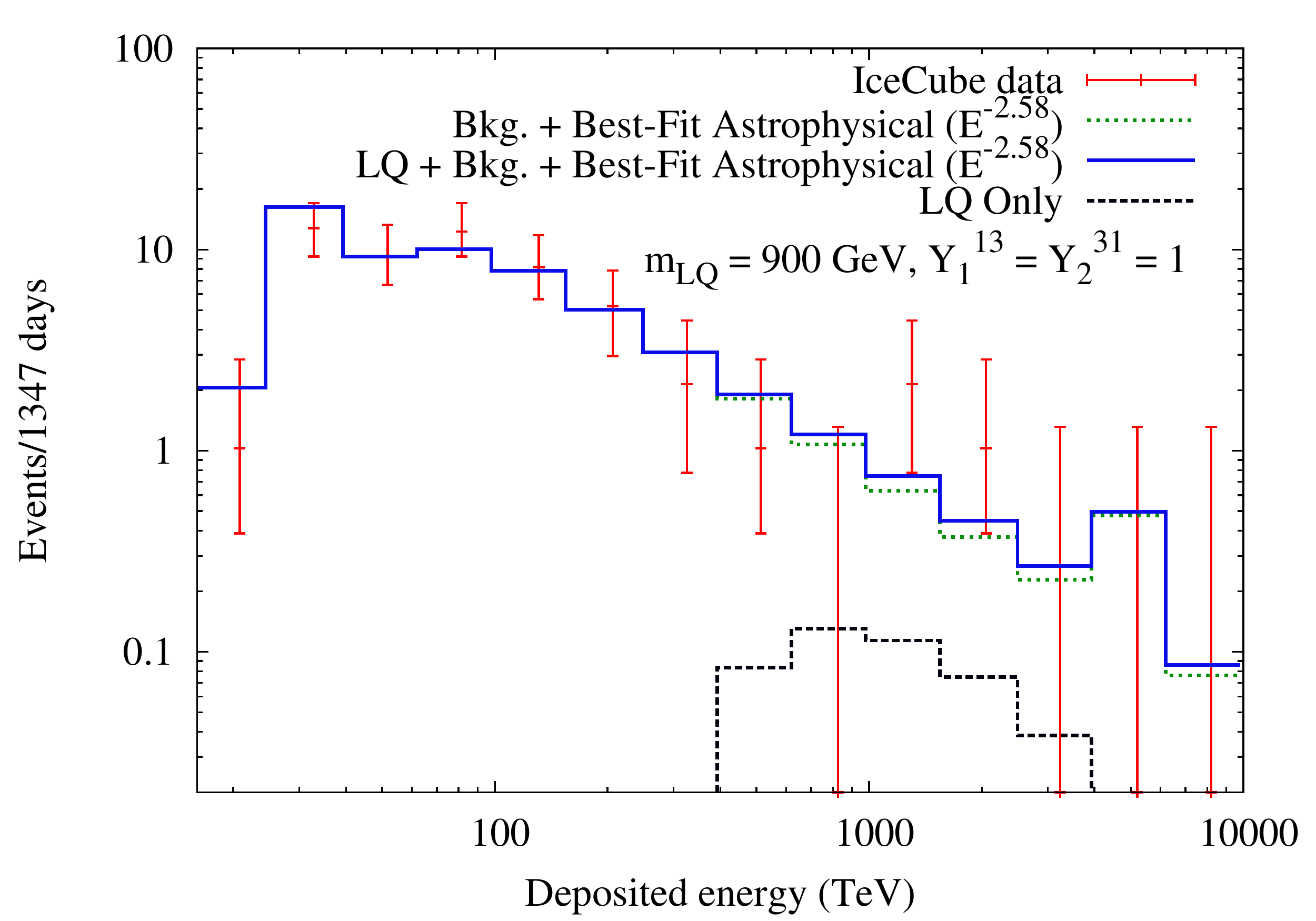}   
   } \\
  \subfloat[\label{f:mlq375}]{
\includegraphics[angle=-90,scale=0.28]{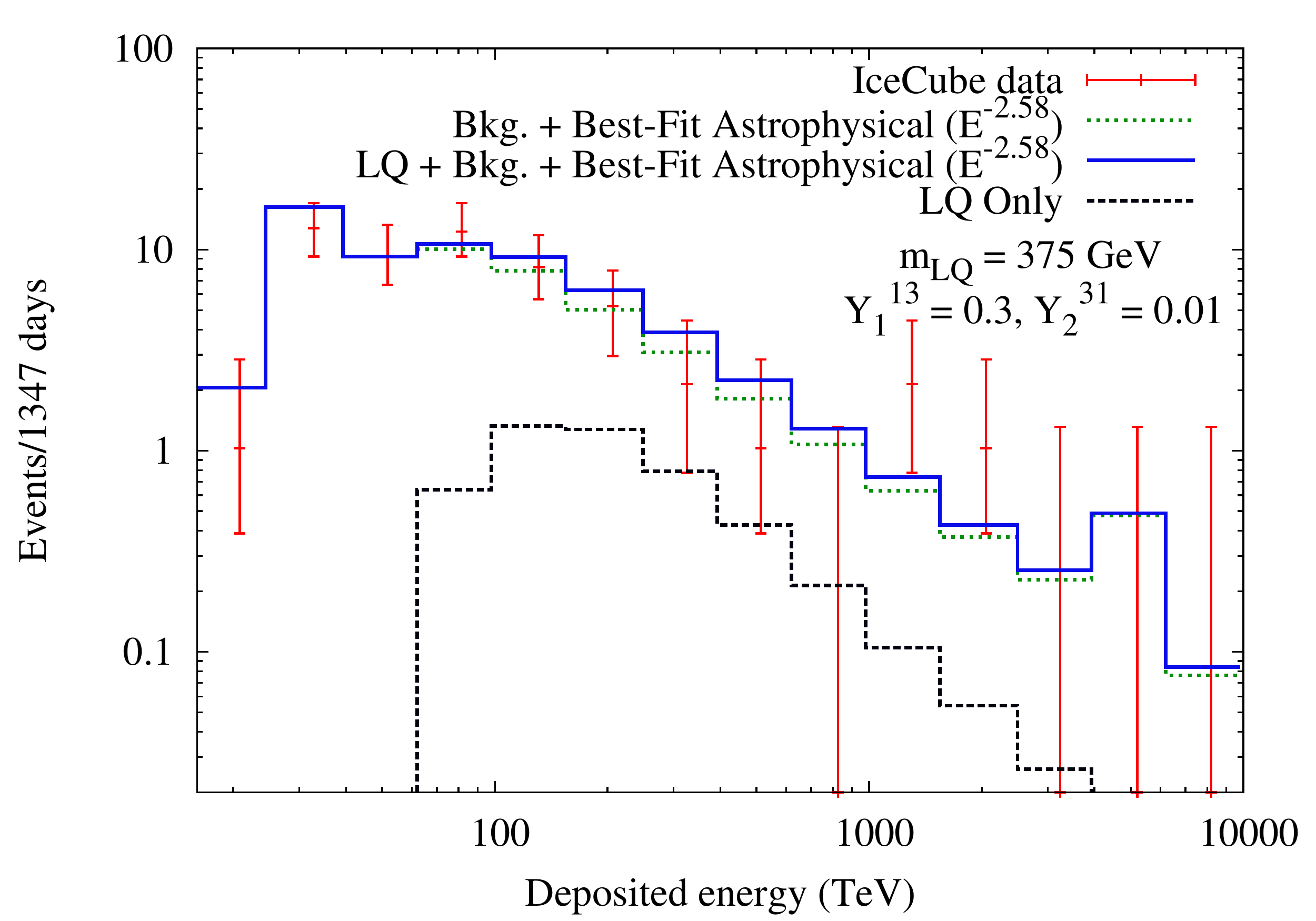}   
   }
  \subfloat[\label{f:mlq1000}]{
\includegraphics[angle=-90,scale=0.28]{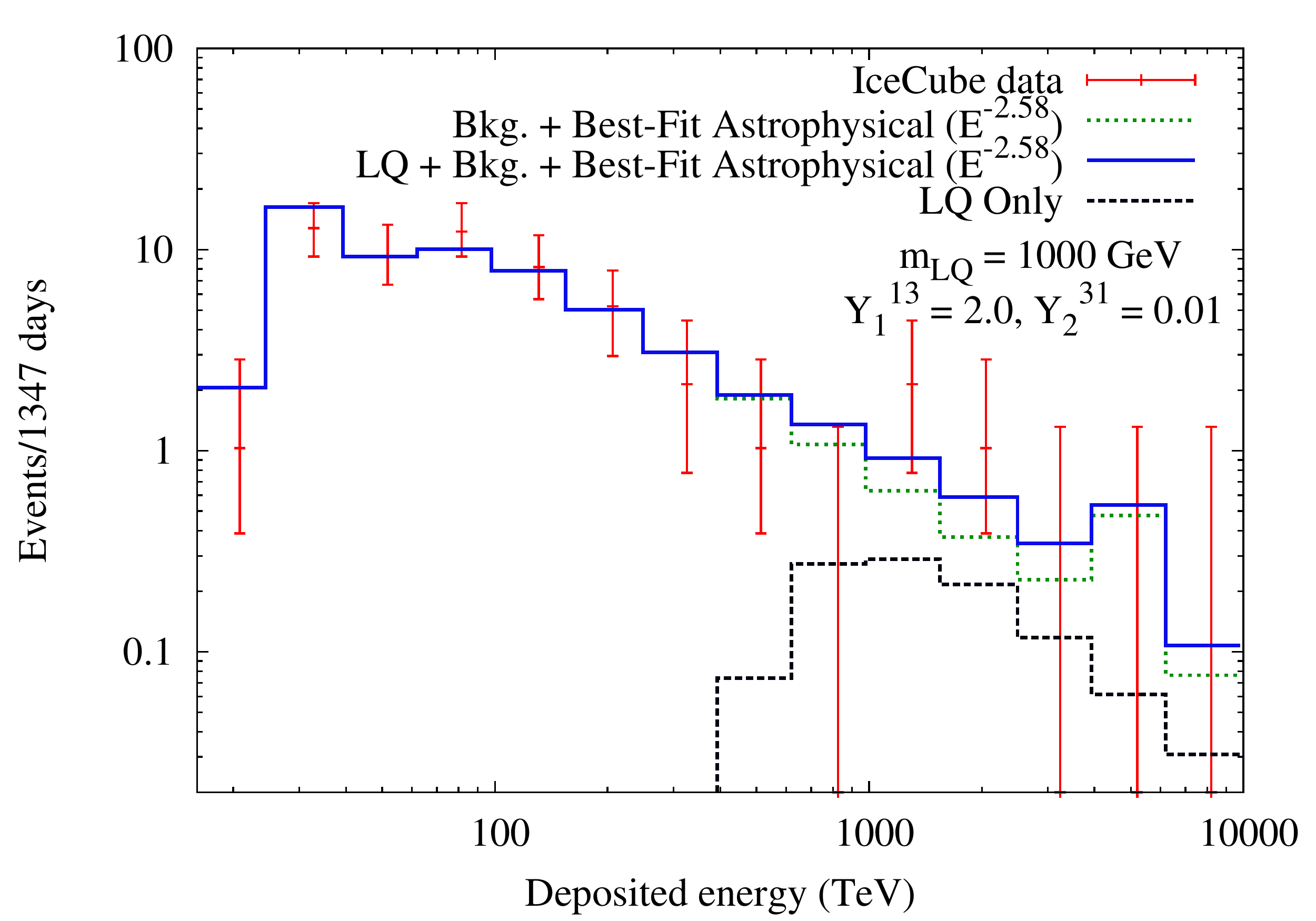}   
   } 
\caption[]{Event distribution at the IceCube with LQ contribution and its comparison with 1347 days data. The panel (a) and (b) shows the event spectra for $m_{LQ} = 780$ GeV and 900 GeV respectively with the relevant couplings taken to be unity. The panel (c) is for $m_{LQ} = 375$ GeV, $Y_{1}^{13} = 0.3$, $Y_{2}^{31} = 0.01$ whereas the panel (d) is for $m_{LQ} = 1000$ GeV, $Y_{1}^{13} = 2.0$, $Y_{2}^{31} = 0.01$.}
\label{fig:evtdistr}    
\end{figure}

For illustrative purposes we choose a few benchmark values of the parameters and show the contribution of LQ on the IceCube events. In Fig.~\ref{fig:evtdistr} we show the event distribution with the LQ contribution for various choice of parameters. For example, in Fig.~\ref{f:mlq780} and \ref{f:mlq900} we show the event distribution for LQ masses 780 GeV and 900 GeV respectively with relevant couplings ($Y_{1}^{13}$ and $Y_{2}^{31}$) taken as unity. Also in Fig.~\ref{f:mlq375} we show the distribution for the leptoquark mass 375 GeV and couplings $Y_{1}^{13} = 0.3$, $Y_{2}^{31} = 0.01$. Fig.~\ref{f:mlq1000} shows the same for $m_{LQ} = 1000$ GeV, $Y_{1}^{13} = 2.0$, $Y_{2}^{31} = 0.01$. From these figures it can be seen that for the unit couplings and LQ mass between 780-900 GeV the LQ contribution helps to better fit the IceCube data in the PeV region. The LQs with masses other than this range can also fit the data by deviating the coupling from unity.

\section{Conclusion}
\label{sec:concl}
The colour-triplet scalar leptoquarks can be utilized to explain the 750 GeV diphoton resonance, recently reported by ATLAS and CMS. Assuming the new resonance to be a singlet scalar $\Phi$ which have couplings with leptoquarks we showed that the observed large cross section $\sigma_{pp\to \gamma \gamma}$ can be accommodated for a range of leptoquark mass and natural coupling between leptoquarks and $\Phi$. The leptoquark loop-induced gluon fusion will produce the scalar $\Phi$ which will subsequently be decayed to $\gamma \gamma$ via the similar leptoquark loop. On the other hand, leptoquarks can be produced via highly energetic neutrino interacting with quarks. A natural place to occur such processes is at the south-pole based neutrino detector, IceCube. It has been shown that the leptoquarks can account for the PeV energy IceCube events. In this work we construct a unified framework to explain both the 750 GeV diphoton resonance and PeV IceCube events by using the scenario of scalar leptoquarks. We consider a general set of scalar leptoquarks that have left-handed couplings with the SM fermions. Below we spell out the main observations.
\begin{itemize}
\item The relevant scalar leptoquarks having left-handed couplings are $S_{1}, S_{3}, R_{2}$ and $\widetilde{R}_{2}$. See Table~\ref{tab1} for their corresponding quantum numbers. The best-suited among these is the $R_{2}$ to explain the large cross section $\sigma_{pp\to \gamma \gamma}$. For the mass range 375 GeV-1 TeV and a natural coupling with the scalar $\Phi$, the leptoquark $R_{2}$ can take care of the large $\sigma_{pp\to \gamma \gamma}$. 
\item The leptoquark $R_{2}$ can be produced resonantly in the interaction of highly energetic incoming neutrinos and quarks at the IceCube. This contributes in the UHE PeV neutrino events at IceCube.   
\item With unit coupling of $R_{2}$ with SM fermions, the PeV events can be explained (within errors) if the mass of $R_{2}$, $m_{LQ}$ lies between 780-900 GeV. This mass range can alter without much adjustment in the relevant couplings. 
\end{itemize} 
Since leptoquarks connecting the first generation quarks to third generation leptons have less stringent bounds on their masses, and these are the one which have been considered in this paper, a specific prediction of this setup is the appearance of $\tau$ in the IceCube, which may be identified by their characteristic double bang signature~\cite{Learned1995}.

%
\bibliographystyle{JHEP}
\bibliography{diphotonRef}
\end{document}